\documentclass[sort&compress]{jpconf}
\usepackage[a4paper]{geometry}
\usepackage{amsmath}
\usepackage{amssymb}
\usepackage{graphicx}
\usepackage[numbers]{natbib}
\usepackage[colorlinks=true,linkcolor=blue,citecolor=blue]
 {hyperref}


\begin{document}
\global\long\def\Ni{n_{\mathrm{i}}}
\global\long\def\Nn{n_{\mathrm{n}}}
\global\long\def\Ti{T_{\mathrm{i}}}
\global\long\def\Rn{R_{\mathrm{n}}}
\global\long\def\Bn{B_{\mathrm{n}}}
\global\long\def\V{V_{\mathrm{i}\zeta}}
\global\long\def\pin{\boldsymbol{\pi}_{\mathrm{n}}}
\global\long\def\mi{m_{\mathrm{i}}}
\global\long\def\Fi{f_{\mathrm{i}}}
\global\long\def\Fgc{f_{\mathrm{i,gc}}}
\global\long\def\Fn{f_{\mathrm{n}}}
\global\long\def\vt{v_{\mathrm{T}}}
\global\long\def\hz{\hat{\boldsymbol{\zeta}}}
\global\long\def\bv{\boldsymbol{v}}
\global\long\def\psin{\psi_{\mathrm{N}}}
\global\long\def\tn{\theta_{\mathrm{n}}}

\title{Edge rotation from momentum transport by neutrals}

\author{JT Omotani$^{1}$, SL Newton$^{1,2}$, I Pusztai$^{1}$ and T F\"ul\"op$^{1}$}

\address{$^{1}$Department of Physics, Chalmers University of Technology,
41296 Gothenburg, Sweden}

\address{$^{2}$CCFE, Culham Science Centre, Abingdon, Oxon, OX14
3DB, UK}

\ead{omotani@chalmers.se}
\begin{abstract}
  Due to their high cross field mobility, neutral atoms can have a
  strong effect on transport even at the low relative densities found
  inside the separatrix. We use a charge-exchange dominated model for
  the neutrals, coupled to neoclassical ions, to calculate momentum
  transport when it is dominated by the neutrals. We can then
  calculate self-consistently the radial electric field and predict
  the intrinsic rotation in an otherwise torque-free plasma. Using a
  numerical solver for the ion distribution to allow arbitrary
  collisionality, 
  we investigate the effects of inverse aspect ratio and
  elongation on plasma rotation. 
  We also
  calculate the rotation of a trace carbon
  impurity, to facilitate future comparison to experiments using
  charge exchange recombination spectroscopy diagnostics.
\end{abstract}


\section{Introduction}


Tokamak plasmas typically rotate, even in the absence of external
momentum input. Rotation is crucial to performance since it stabilizes
magnetohydrodynamic instabilities such as the resistive wall mode
\citep{Hender2007} and flow shear suppresses turbulence in the
pedestal, leading to the high-confinement mode (H-mode) \citep{Terry00}.
Intrinsic rotation is particularly important for future tokamaks such
as ITER because the torque from NBI heating will be weaker in
larger devices. Neutral particles, which are always present in the
edge of the confined plasma volume, provide a mechanism to generate
intrinsic rotation. They may give large momentum transport even at low
relative density due to their high cross field mobility, as has been
demonstrated theoretically
\citep{Hazeltine1992,Catto1994,Helander1994,Catto98,Fulop98_1,Fulop98_2,Fulop01,Fulop02,Helander2003}.
The importance of neutrals for the edge plasma is confirmed by
experimental evidence of the influence of neutral density and location
on H-mode confinement and the low to high confinement transition
threshold
\citep{Carreras98,Owen98,Gohil01,Boivin00,Field02_MAST,Valovic02_COMPASS,Fukuda00,Field2004,Maingi2004,Joffrin14_JET,Tamain2015}.
The neutrals may be localized poloidally due to gas fuelling at a
particular location, or divertor recycling causing the neutrals to be
concentrated near the X-point. The resulting momentum transport and
rotation is then controlled by the location of the neutrals
\citep{Fulop02,Helander2003,Omotani2016}. In Ref.~\citep{Omotani2016}
we presented a numerical framework with which the rotation can be
calculated for such scenarios, allowing for arbitrary plasma
collisionality.  In this paper we validate our numerical procedure
against the analytical limits given in
Refs.~\citep{Fulop02,Helander2003} for asymptotic collisionality
regimes. We investigate the effects of the lowest order equilibrium
shaping parameters (i.e. those with the smallest poloidal mode
numbers), inverse aspect ratio $\epsilon$ and elongation $\kappa$.
Finally, we consider impurity rotation. Plasma rotation is usually
determined by charge exchange recombination spectroscopy
\citep{Isler1994} which often measures the rotation of an impurity
species. The main ion and impurity rotation are known to differ
\citep{Kim1991}. We demonstrate that by including a trace impurity
species in our simulations, both the bulk ion and the impurity
rotation can be calculated within our framework for arbitrary
collisionality of either species.

\section{Momentum transport by neutrals} \subsection{Model} The
evolution of the radial electric field $E_{r}$ in a tokamak plasma is
determined by the radial current through Amp\`ere's law. In steady
state the radial current must vanish. 
The radial particle fluxes $\boldsymbol{\Gamma}_a$ of the charged
species are then determined by the toroidal component of the momentum
equations for species $a$
\citep{Helander_book},
\begin{align}
  \left\langle
  e_{a}\boldsymbol{\Gamma}_{a}\cdot\nabla\psi\right\rangle &
  =\left\langle
  R\hat{\boldsymbol{\zeta}}\cdot\nabla\cdot\boldsymbol{\pi}_{a}\right\rangle
  -\left\langle RF_{a\zeta}\right\rangle -\left\langle
  n_{a}e_{a}RE_{\zeta}^{(A)}\right\rangle \label{eq:toroidal-mtm}
\end{align}
for ions, electrons or neutrals. Here $\left\langle...\right\rangle$
denotes the flux surface average, $2\pi\psi$ is the poloidal flux, $R$
is the major radius,
$\hat{\boldsymbol{\zeta}}=\nabla\zeta/\left|\nabla\zeta\right|$ is the
unit vector in the direction of the toroidal angle $\zeta$,
$\boldsymbol{E}^{(A)}$ is the induced electric field and for the
species $a$, $e_a$ is the charge, $\boldsymbol{\pi}_a$ the viscosity
tensor and $\boldsymbol{F}_{a}=\sum_{b}\boldsymbol{F}_{ab}$ the total
friction force due to collisions with all other species $b$.
$E_{\zeta}^{(A)}$ does not contribute to the radial current due to
quasineutrality and $\sum_{a}\boldsymbol{F}_{a}=0$ as momentum is
conserved in collisions. For charged particles the viscosity component
in \eqref{eq:toroidal-mtm} is typically negligible
\citep{Helander_book}, leading to automatic ambipolarity at lowest
order in the gyroradius expansion of standard neoclassical theory for
a fully ionized plasma, so the radial electric field
$E_{r}=-\left|\nabla\psi\right|\partial\Phi_0/\partial\psi$ is not
constrained at this order
\citep{Rosenbluth1971,Hazeltine1974,Catto2005}. However, neutrals have
a non-negligible viscosity due to their high radial mobility which, as
has been shown in the previous analytical work and is outlined below,
depends on the toroidal velocity directly and therefore can constrain
the electric field.

Since $\nabla(R\hat{\boldsymbol{\zeta}})$ is an antisymmetric tensor,
while $\boldsymbol{\pi}_{a}$ is symmetric, then in the absence of
external torques the vanishing of the radial current in steady state,
where the sum on species $a$ is over ions, electrons and neutrals,
implies
\begin{equation} 0
=j_{r}=\sum_{a}\left\langle
e_{a}\boldsymbol{\Gamma}_{a}\cdot\nabla\psi\right\rangle \nonumber
\approx\left\langle \nabla\cdot\left(R\hz\cdot\pin\right)\right\rangle
=\frac{1}{V'}\frac{d}{d\psi}\left(V'\left\langle
R\hz\cdot\pin\cdot\nabla\psi\right\rangle \right)
\end{equation}
\begin{equation} \Rightarrow \left\langle
R\hz\cdot\pin\cdot\nabla\psi\right\rangle =0,\label{eq:constraint}
\end{equation}
with $V'=\oint
d\theta\,\left(\boldsymbol{B}\cdot\nabla\theta\right)^{-1}$, and we
have noted that in the absence of external torque the boundary
condition in the interior of the plasma is that the radial flux
of toroidal angular momentum vanishes.

To compute the viscosity $\pin$ of the neutrals (denoted by subscript
n) we solve the steady state neutral kinetic equation \citep{Catto98}, assuming
charge exchange (CX) dominates so that we may neglect other
atomic processes,
\begin{align} \boldsymbol{v}\cdot\nabla\Fn &
=\frac{1}{\tau_\mathrm{CX}}\left(\frac{\Nn}{\Ni}\Fi-\Fn\right),\label{eq:neutral-kinetic-equation}
\end{align}
where $f_a$ and $n_a$ are the distribution function and density of a
species $a$ (we denote the bulk ions by subscript i).  The CX
collision frequency is $\tau_\mathrm{CX}^{-1}=\Ni\left\langle \sigma
v\right\rangle
_{\mathrm{CX}}\simeq2.93\Ni\sigma_{\mathrm{CX}}\left(\Ti/\mi\right)^{1/2}$
\citep{Catto98}, which is larger than the ionization or recombination
rates for tokamak edge parameters \citep{Helander1999}; the CX rate is
$\sim 40\%$ larger than the ionization rate at the $300\;\mathrm{eV}$
temperature we consider below, while recombination is negligible above
a few eV \citep{ADAS}. Since we
assume that the neutrals are dominated by charge exchange collisions
with the ions, their temperature is the same as the ion temperature to
lowest order, as the solution below shows.  $\left\langle\sigma
v\right\rangle_\mathrm{CX}$ is the thermal charge exchange rate
coefficient and $\sigma_\mathrm{CX}$ is the charge exchange cross
section for thermal particles. We will assume that the relative
neutral density $\Nn/\Ni$ is small enough that the ion distribution is
not affected directly by the neutrals, i.e.~$\Fi$ is independent of
$\Fn$. We can then solve \eqref{eq:neutral-kinetic-equation} with a
short mean-free-path expansion $ \Fn  =\Fn^{(0)}+\Fn^{(1)}+...  $ in
$\tau_\mathrm{CX}\vt/L\ll1$, where $\vt=\sqrt{2\Ti/\mi}$ is the
thermal velocity, with the ion temperature $\Ti$ and mass $\mi$, and
$L$ is a typical length scale of plasma parameters, giving
\begin{equation}
\Fn^{(0)} =\frac{\Nn}{\Ni}\Fi;\quad
\Fn^{(1)} =-\tau_\mathrm{CX}\boldsymbol{v}\cdot\nabla\Fn^{(0)}=-\tau_\mathrm{CX}\boldsymbol{v}\cdot\nabla\left(\frac{\Nn}{\Ni}\Fi\right).
\end{equation}
The neutral viscosity is then
\begin{equation}
\pin =\mi\int
  d^{3}v\,\left(\boldsymbol{v}\boldsymbol{v}-\frac{1}{3}v^{2}\mathbb{I}\right)\Fn
=\frac{\Nn}{\Ni}\boldsymbol{\pi}_{\mathrm{i}}-\tau_\mathrm{CX}\mi\int d^{3}v\,\left(\boldsymbol{v}\boldsymbol{v}-\frac{1}{3}v^{2}\mathbb{I}\right)\boldsymbol{v}\cdot\nabla\left(\frac{\Nn}{\Ni}\Fi\right).
\end{equation}
The first term can be neglected as the off-diagonal components of
$\boldsymbol{\pi}_{\mathrm{i}}$ are small, leaving
\begin{align}
R\hz\cdot\pin\cdot\nabla\psi &\approx-R\tau_\mathrm{CX}\mi\int
d^{3}v\,\left(\hz\cdot\bv\right)\left(\nabla\psi\cdot\bv\right)\bv\cdot\nabla\left(\frac{\Nn}{\Ni}\Fi\right)
\nonumber \\
&\approx-\frac{R\tau_\mathrm{CX}\mi}{\Ni}\frac{d\Nn}{d\psi}\int
d^{3}v\,\left(\hz\cdot\bv\right)\left(\nabla\psi\cdot\bv\right)^{2}\Fi,\label{eq:pin}
\end{align}
assuming that the radial gradient of $\Nn$ dominates over those of the
background plasma profiles and over any poloidal gradients.

Accounting for the gyroradius correction, the ion distribution function at
the particle position $\boldsymbol{r}$ is
\begin{align}
\Fi(\boldsymbol{r}) & \approx
  f_{\mathrm{i,gc}0}(\boldsymbol{r})-\frac{e\Phi_{1}}{\Ti}f_{\mathrm{i,gc}0}(\boldsymbol{r})-\boldsymbol{\rho}\cdot\nabla
  f_{\mathrm{i,gc}0}(\boldsymbol{r})+g_{\mathrm{i}}(\boldsymbol{r}),\label{eq:fi}
\end{align}
where $f_{\mathrm{i,gc}0}$ is a
Maxwellian, $\boldsymbol{\rho}=\boldsymbol{r}-\boldsymbol{R}_\mathrm{gc}$
is the gyroradius vector with $\boldsymbol{R}_\mathrm{gc}$ the guiding
centre position, $\Phi_{1}=\Phi-\left\langle \Phi\right\rangle $ is
the poloidally varying part of the electrostatic potential $\Phi$ and
0, 1 subscripts refer to the order in a $\delta=\rho/L$ gyroradius
expansion.  We will obtain the non-adiabatic part of the perturbed
distribution function,
$g_{\mathrm{i}}=f_{\mathrm{i,gc}1}+\frac{e_{\mathrm{i}}\Phi_{1}}{\Ti}f_{\mathrm{i,gc}0}$,
from numerical solutions of the first order drift kinetic equation.
The first two terms on the right of (\ref{eq:fi}) are isotropic in
velocity, so do not contribute to (\ref{eq:pin}), and we have
\begin{align}
R\hz\cdot\pin\cdot\nabla\psi &
  \approx\left(R\hz\cdot\pin\cdot\nabla\psi\right)^{\mathrm{(dia)}}+\left(R\hz\cdot\pin\cdot\nabla\psi\right)^{\mathrm{(neo)}},
  \label{eq:pin-tot}
\end{align}
where 
\begin{align}
\left(R\hz\cdot\pin\cdot\nabla\psi\right)^{\mathrm{(dia)}} &
=\frac{\tau_\mathrm{CX}
R^{4}B_{\mathrm{p}}^{4}\Ti^{2}}{eB^{2}}\frac{d\Nn}{d\psi}\left(\frac{d\ln
p_{\mathrm{i}}}{d\psi}+\frac{e}{\Ti}\frac{d\Phi_{0}}{d\psi}+\frac{d\ln\Ti}{d\psi}\right)\label{eq:pi-dia}\\
\left(R\hz\cdot\pin\cdot\nabla\psi\right)^{\mathrm{(neo)}} &
=-\frac{\tau_\mathrm{CX}
R^{2}B_{\mathrm{p}}^{2}I\mi}{2B\Ni}\frac{d\Nn}{d\psi}\int
d^{3}v\,v_{\|}v_{\perp}^{2}g_{\mathrm{i}},\label{eq:pi-neo}
\end{align}
with $B_\mathrm{p}$ the poloidal magnetic field strength,
$I=R\boldsymbol{B}\cdot\hat{\boldsymbol\zeta}$ and parallel and
perpendicular taken relative to the background magnetic field.  Drift
velocity terms do not appear in \eqref{eq:pi-neo} as they are second
order in $\delta$. This momentum flux is first order in $\delta$, and
therefore large compared to that of the ions, as the neutral
distribution $\Fn^{(1)}$ is strongly anisotropic in gyroangle.

\subsection{Interpretation\label{sub:Interpretation}}

It is useful to write the momentum flux \eqref{eq:pin-tot} in terms of
the toroidal velocity and heat flux in order to explain the physical
origins of the intrinsic momentum flux.  This is done as follows. We
can write (\ref{eq:pi-dia}) in terms of the toroidal components of the
diamagnetic flow and heat flux \citep{Helander2003}
\begin{equation}
  V_{\zeta}^{\mathrm{(dia)}} =-\frac{RB_{\mathrm{p}}^{2}\Ti}{eB^{2}}\left(\frac{d\ln
p_{\mathrm{i}}}{d\psi}+\frac{e}{\Ti}\frac{d\Phi_{0}}{d\psi}\right);\quad
\frac{2q_{\zeta}^{\mathrm{(dia)}}}{5p_{\mathrm{i}}}
=-\frac{RB_{\mathrm{p}}^{2}\Ti}{eB^{2}}\frac{d\ln\Ti}{d\psi},
\end{equation}
as
\begin{align}
\left(R\hz\cdot\pin\cdot\nabla\psi\right)^{\mathrm{(dia)}} &
  =-\tau_\mathrm{CX}
  R^{3}B_{\mathrm{p}}^{2}\Ti\frac{d\Nn}{d\psi}\left(V_{\zeta}^\mathrm{(dia)}+\frac{2q_{\zeta}^\mathrm{(dia)}}{5p_{\mathrm{i}}}\right).
\end{align}
To rewrite \eqref{eq:pi-neo} note that
$v_{\|}v_{\perp}^{2}=\frac{2}{5}v_{\|}v^{2}-\frac{2}{5}v^{3}P_{3}\left(\xi\right)$
where $\xi=v_{\|}/v$ is the cosine of the pitch angle and
$P_{3}(\xi)\equiv5\xi^{3}/2-3\xi/2$ is the third order Legendre
polynomial. The neoclassical part of the total heat flux is 
\begin{align}
\frac{I}{RB}\int
d^{3}v\,\frac{\mi}{2}v_{\|}v^{2}g_{\mathrm{i}}\equiv\frac{I}{RB}Q_{\mathrm{i}\|}^{\mathrm{(neo)}}
& \approx Q_{\mathrm{i}\zeta}^{\mathrm{(neo)}} 
=q_{\mathrm{i}\zeta}^{\mathrm{(neo)}}+\frac{5p_{\mathrm{i}}}{2}V_{\mathrm{i}\zeta}^{\mathrm{(neo)}}+\mathcal{O}(\delta^{3})
\end{align}
and so (\ref{eq:pi-neo}) gives
\begin{align}
\left(R\hz\cdot\pin\cdot\nabla\psi\right)^{\mathrm{(neo)}} &
\approx-\tau_\mathrm{CX}
R^{3}B_{\mathrm{p}}^{2}\Ti\frac{d\Nn}{d\psi}\left(V_{\mathrm{i}\zeta}^{\mathrm{(neo)}}+\frac{2}{5p_{\mathrm{i}}}q_{\mathrm{i}\zeta}^{\mathrm{(neo)}}\right)+\frac{\tau_\mathrm{CX}
R^{2}B_{\mathrm{p}}^{2}I\mi}{5B\Ni}\frac{d\Nn}{d\psi}\int
d^{3}v\,v^{3}P_{3}(\xi)g_{\mathrm{i}}.
\end{align}
Thus we have finally
\begin{align}
\left(R\hz\cdot\pin\cdot\nabla\psi\right) & \approx-\tau_\mathrm{CX}
  R^{3}B_{\mathrm{p}}^{2}\Ti\frac{d\Nn}{d\psi}\left(V_{\mathrm{i}\zeta}+\frac{2}{5p_{\mathrm{i}}}q_{\mathrm{i}\zeta}\right)+\frac{\tau_\mathrm{CX}
  R^{2}B_{\mathrm{p}}^{2}I\mi}{5B\Ni}\frac{d\Nn}{d\psi}\int
  d^{3}v\,v^{3}P_{3}(\xi)g_{\mathrm{i}} \label{eq:approx-flux}
\end{align}
and we see that the steady state condition \eqref{eq:constraint} is
largely set by a balance between the toroidal friction and heat
friction, with an additional contribution from the third Legendre component of
the distribution function.\footnote{Note that this form is evaluated
for arbitrary bulk ion collisionality. The $g_\mathrm{i}$ appearing
here corresponds to the usual $F+g^\mathrm{banana}$ in the low
collisionality regime, giving rise to the equivalent expression (17)
in Ref.~\citep{Helander2003}.}

The plasma flow and heat flux take the general form \citep{Helander_book}
\begin{align}
\boldsymbol{V}_{\mathrm{i}} &
=-\left(\frac{d\Phi_{0}}{d\psi}+\frac{1}{e\Ni}\frac{dp_{\mathrm{i}}}{d\psi}\right)R\hz-\frac{kI}{e\left\langle B^{2}\right\rangle }\frac{d\Ti}{d\psi}\boldsymbol{B}\label{eq:V}\\
\boldsymbol{q}_{\mathrm{i}} &
=-\frac{5p_{\mathrm{i}}}{2e}\frac{d\Ti}{d\psi}R\hz+\frac{5lIp_{\mathrm{i}}}{2e\left\langle
B^{2}\right\rangle }\frac{d\Ti}{d\psi}\boldsymbol{B},\label{eq:q}
\end{align}
where $k$ and $l$ are dimensionless parameters that depend on the
plasma collisionality. $k$ varies from positive in the
Pfirsch-Schl\"uter regime, (\ref{eq:k_PS}), to negative in the banana
regime, (\ref{eq:kl_banana}), while $l$ remains positive, with $l=1$
in the Pfirsch-Schl\"uter regime and slightly smaller in the banana
regime, (\ref{eq:kl_banana}).

The neutral source is often poloidally localized, for example from a
gas-puff valve or divertor recycling. Modelling the neutral density as
a $\delta$-function at $\theta=\tn$, the flux surface average in
(\ref{eq:constraint}) simply evaluates its argument at the poloidal
location $\tn$. Neglecting here, for illustrative purposes, the
$P_3$ term we can use (\ref{eq:approx-flux}) in (\ref{eq:constraint})
to solve for the electric field, and hence also the rotation, as
\begin{align}
-\frac{d\Phi_{0}}{d\psi} &
\approx\frac{\Ti}{e\Ni}\frac{d\Ni}{d\psi}+\frac{1}{e}\left(2+\frac{(k-l)I^{2}}{\Rn^{2}\left\langle
B^{2}\right\rangle }\right)\frac{d\Ti}{d\psi};\quad
\V\approx\left(R
  +\frac{(k-l)I^2R}{\Rn^2\left\langle B^2\right\rangle}
  -\frac{kI^2}{R\left\langle B^2\right\rangle}
  \right)\frac{1}{e}\frac{d\Ti}{d\psi},\label{eq:ErVzeta_approx}
\end{align}
where $\Rn=R(\tn)$ is the major radius at which the neutrals are
located.  Thus the sign of $(k-l)$ determines the sign of the
variation of $E_r$ with $\Rn$.  At high collisionality $(k-l)<0$ so
$E_r$ is more negative for small $\Rn$, while at low collisionality
$(k-l)>0$ so $E_r$ is more positive for small $\Rn$. An accurate
analytic form in the banana regime was given in Ref.
\citep{Helander2003}, (\ref{eq:Er_banana}) here; it has the same
$(k-l)$ prefactor for its $\Rn$ dependence, indicating that the
preceding discussion does capture the origin of the trends.  The
toroidal velocity has a similar variation with $\Rn$ as it comes from
substituting $d\Phi_{0}/d\psi$ into (\ref{eq:V}).  These features can
be identified in the numerical results shown in the following section.

\section{Simulation results}

As can be seen from Eq.~\eqref{eq:pi-neo} we need the ion guiding
centre distribution function to determine the steady state radial
electric field and toroidal flow. To calculate this we use the
Pedestal and Edge Radially-global Fokker-Planck Evaluation of
Collisional Transport (\textsc{perfect}) neoclassical solver
\citep{Landreman14_PERFECT}, used here as a radially-local solver.
Taking a guess for $d\Phi_0/d\psi$ we use the distribution function
from \textsc{perfect} to calculate the neutral momentum flux, and
iterate on $d\Phi_0/d\psi$ to find the value consistent with the
steady state condition, Eq.~\ref{eq:constraint}. While experimental
equilibria can be used, here we take model divertor magnetic
geometries based on analytical solutions to the Grad-Shafranov
equation \citep{Cerfon10}, allowing trends in the rotation behaviour
with shaping parameters to be identified. The geometry is
parameterized in terms of the inverse aspect ratio $\epsilon$,
elongation $\kappa$ and triangularity $\delta$, as well as the major
radius $R_{\mathrm{X}}$ and height $Z_{\mathrm{X}}$ of the X-point,
along with the toroidal $\beta_\mathrm{t}=2 \mu_0 \bar p /B_0^2$
(where $\bar p$ is the volume averaged pressure) and plasma current.
Dimensional scales are set by the major radius of the plasma centre
$R_{0}$ and the vacuum magnetic field at the centre $B_{0}$. We take
the nominal scan parameters representative of ITER \citep{Cerfon10}:
$R_{0}=6.2\,\mathrm{m}$, $B_{0}=5.3\,\mathrm{T}$, $\epsilon=0.32$,
$\kappa=1.7$, $\delta=0.33$,
$R_{\mathrm{X}}=(1-1.1\delta\epsilon)R_{0}$,
$Z_{\mathrm{X}}=-1.1\kappa\epsilon R_{0}$ and $\beta_\mathrm{t}=0.05$.
We allow the plasma current to vary to keep the safety factor $q_{95}$
of the flux surface with normalized poloidal flux
$\psi_\mathrm{N}=0.95$ fixed to the value 2.84, corresponding to a
plasma current of 15~MA in the baseline equilibrium. The neutrals are
poloidally localized, represented by a delta function\footnote{The
delta function representation takes the poloidal width of the neutral
density variation to be narrow compared to the minor radius, but the
poloidal width should be understood to be large compared to the radial
gradient scale length, so that poloidal derivatives may be neglected.}
in poloidal angle
$\Nn=\hat{n}_{\mathrm{n}}(\psi)\delta(\theta-\tn)$, on the
$\psin=0.95$ flux surface.  The main ion species and neutrals are both
deuterium and we do not need to solve the electron kinetic equation
here due to the small mass ratio.

Below we show scans plotted at fixed collisionality. We use the
collisionality $\hat{\nu}=\nu_\mathrm{ii}L_\|/\vt$, with $\nu_\mathrm{ii}$ the
ion-ion collision frequency, calculated using the connection length
$L_\|=\int_{\theta=0}^{\theta=2\pi}dl_\|/2\pi$, where $dl_\|$ is the
line element parallel to $\boldsymbol{B}$, which reduces to
$L_\|\approx qR_0$ in the large aspect ratio limit. The baseline value
is $\nu_\mathrm{ii}L_\|/\vt\approx 1.26$ corresponding to
$\Ti=300\;\mathrm{eV}$ and $\Ni=10^{20}\;\mathrm{m}^{-3}$ in the
baseline equilibrium. We set the collisionality by varying the density
at fixed temperature and logarithmic temperature gradient. To give
explicit rotation velocities and electric fields in physical units, we
choose a scale length for the ion temperature gradient, at the
outboard midplane, of $L_{T\mathrm{i}}=10\;\mathrm{cm}$ in the baseline
equilibrium.  In section \ref{sub:shaping} where we scan the
equilibrium geometry, we hold fixed the gradient in poloidal flux at
the value
$\partial\ln\Ti/\partial\psi=-1/L_{T\mathrm{i}}\left|\nabla\psi\right|\approx0.911\;\mathrm{T^{-1}\;m^{-2}}$
corresponding to $L_{T\mathrm{i}}=10\;\mathrm{cm}$ in the baseline
equilibrium. We set $\partial\Ni/\partial\psi=0$ here. The density
gradient $\partial\Ni/\partial\psi$ and electrostatic potential
gradient $\partial\Phi_0/\partial\psi$ terms in the drift kinetic
equation have identical velocity space structures. Therefore when we
solve for $\partial\Phi_0/\partial\psi$ any density gradient gives an
offset to $\partial\Phi_0/\partial\psi$ without affecting the flow on
the flux surface we consider, as can be seen in
Eq.~\eqref{eq:ErVzeta_approx}.

\subsection{Comparison to analytical theory}

\begin{figure}[tbp]
  \includegraphics[width=0.47\columnwidth]{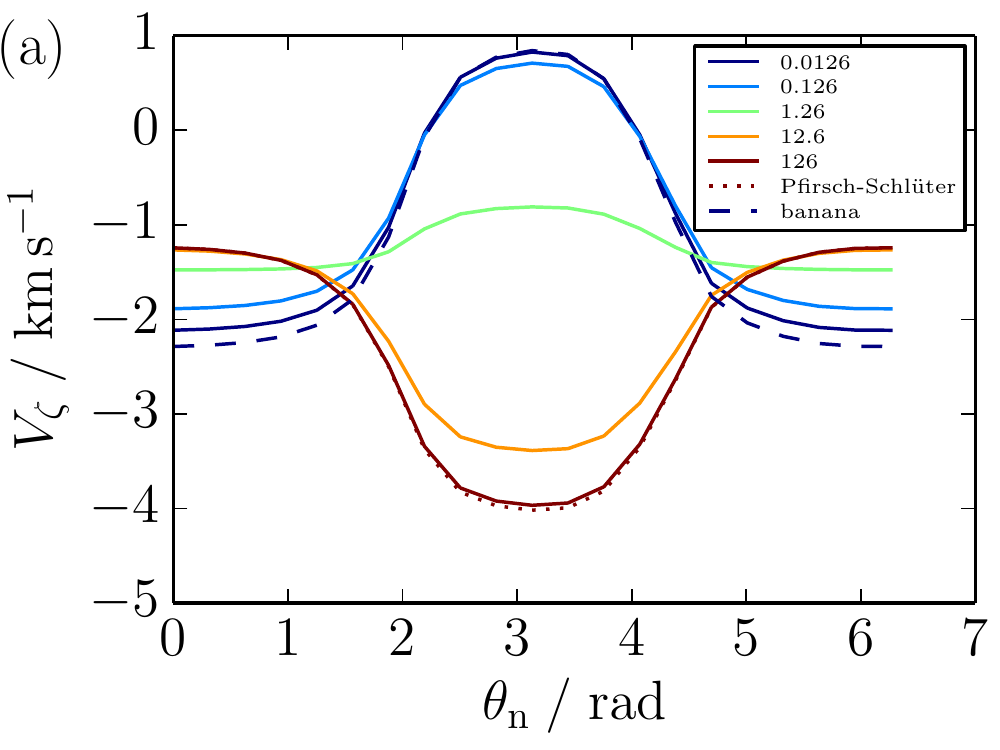}\hfill{}\includegraphics[width=0.47\columnwidth]{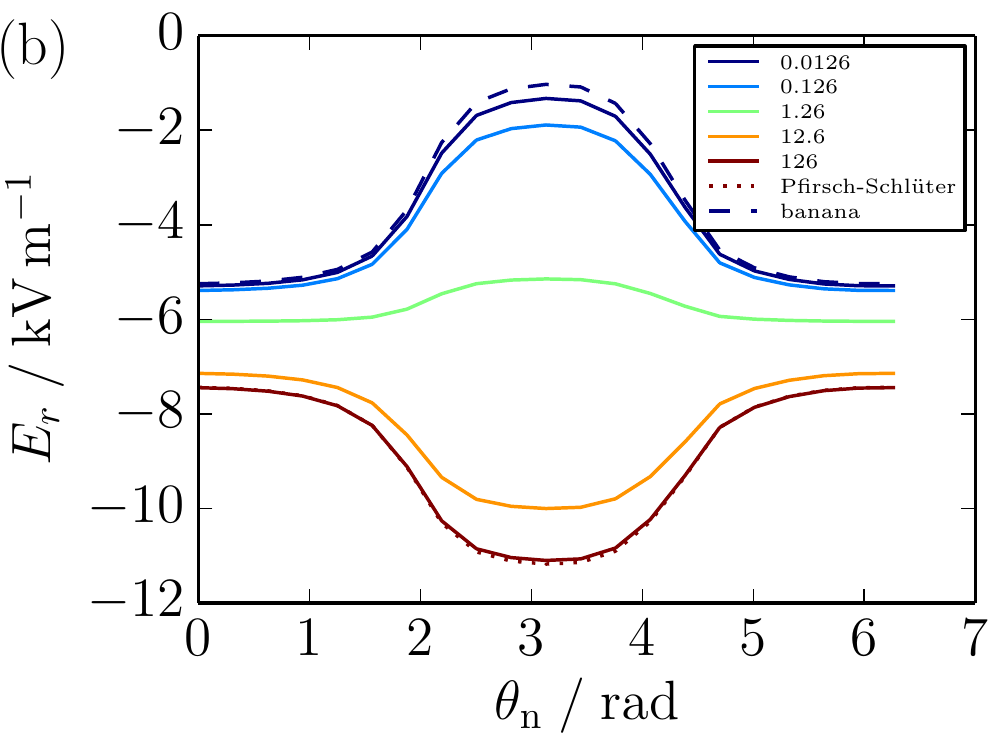}

  \caption{Comparison of analytical and numerical results for the
    dependence of (a) toroidal flow velocity and (b) radial electric
    field at the outboard midplane on poloidal location of the
    neutrals. 
    Dashed lines show the banana regime analytic results,
    (\ref{eq:Vzeta_banana}) and (\ref{eq:Er_banana}), and dotted lines
    the Pfirsch-Schl\"uter analytic results, (\ref{eq:Vzeta_PS}) and
    (\ref{eq:Er_PS}). Solid lines show numerical scans, with colours
    from blue to red corresponding to low to high collisionality
    $\hat\nu$, as shown in the legend.
    A typical ion temperature scale length of 10 cm is taken, to
    remove the normalizations and give explicit rotation velocity and
    electric field values. $\tn=0$ is the outboard midplane.
    \label{fig:analytic-comparison}}
\end{figure}

Analytical forms for the radial electric field and toroidal flow that
result from momentum transport by neutrals were calculated in the
Pfirsch-Schl\"uter and banana regimes in Ref.~\citep{Helander2003}.
The electric field $-d\Phi_0/d\psi$ is a flux function, but $\V$ varies
poloidally and depends on both the major radius where the neutrals are
localized, $\Rn$, and the major radius where $\V$ is evaluated, $R$.
In the Pfirsch-Schl\"uter regime
\begin{align}
  -\frac{d\Phi_0}{d\psi}(\Rn) &
  =\frac{\Ti}{e\Ni}\frac{d\Ni}{d\psi}+\frac{1}{e}\frac{d\Ti}{d\psi}\left(2+\frac{(k-1)I^{2}}{\left\langle B^{2}\right\rangle \Rn}\right)\label{eq:Er_PS}\\
\V(R,\Rn) & =\frac{I^{2}}{e\left\langle B^{2}\right\rangle R}\frac{d\Ti}{d\psi}\left(\frac{\left\langle B^{2}\right\rangle R^{2}}{I^{2}}-\frac{R^{2}}{\Rn^{2}}+k\left(\frac{R^{2}}{\Rn^{2}}-1\right)\right)\label{eq:Vzeta_PS}
\end{align}
and the coefficient $k$ is
\begin{align}
k & =1.97+0.09\frac{\left\langle B^{2}\right\rangle \left\langle
\left(\nabla_{\|}\ln B\right)^{2}\right\rangle }{\left\langle
\left(\nabla_{\|}B\right)^{2}\right\rangle },\label{eq:k_PS}
\end{align}
with the numerical prefactors corrected by a more accurate, numerical
solution to the relevant Spitzer problems \citep{LandremanPrivate}.

In the banana regime,
\begin{align}
-\frac{d\Phi_0}{d\psi}(\Rn)
&=\frac{\Ti}{e\Ni}\frac{d\Ni}{d\psi}+\frac{1}{e}\frac{d\Ti}{d\psi}\left(2+\left(k-l\right)\frac{f_{2}}{f_{c}}\frac{\Bn I^{2}}{\left\langle B^{2}\right\rangle ^{3/2}\Rn^{2}}\right)\label{eq:Er_banana}\\
\V(R,\Rn) & =\frac{I^{2}}{e\left\langle B^{2}\right\rangle
R}\frac{d\Ti}{d\psi}\left(\frac{\left\langle B^{2}\right\rangle
R^{2}}{I^{2}}-k+(k-l)\frac{f_{2}}{f_{\mathrm{c}}}\frac{R^{2}\Bn}{\Rn^{2}\sqrt{\left\langle
B^{2}\right\rangle }}\right),\label{eq:Vzeta_banana}
\end{align}
where for a pure plasma
\begin{equation}
k =-1.17/(1+0.46f_{\mathrm{t}}/f_{\mathrm{c}}); \quad
l =1/(1+0.46f_{\mathrm{t}}/f_{\mathrm{c}}), \label{eq:kl_banana}
\end{equation}
using an interpolation formula \citep{Helander_book,Taguchi1988}
between the exact results for large and unit aspect ratio, with the
usual definition of the fraction of circulating particles
\begin{equation}
f_{\mathrm{c}} =1-f_{\mathrm{t}}=\frac{3 }{4}\left\langle
B^{2}\right\rangle\int_{0}^{B_{\mathrm{max}}^{-1}}d\lambda\,\frac{\lambda}{\left\langle
\sqrt{1-\lambda B}\right\rangle }; \quad
f_{2} =\frac{15}{16}\left\langle B^{2}\right\rangle ^{3/2}\int_{0}^{B_{\mathrm{max}}^{-1}}d\lambda\,\frac{\lambda^{2}}{\left\langle \sqrt{1-\lambda B}\right\rangle }.
\end{equation}

Fig.~\ref{fig:analytic-comparison} shows the simulation results of a
scan in the poloidal angle $\tn$ at which the neutrals are
localized in the nominal geometry for a wide range of
collisionalities. The analytical limits quoted above are also shown
and good agreement can be seen at both high and low collisionality,
verifying our numerical implementation. The trends are as expected
from the discussion in Sec. \ref{sub:Interpretation}.


\subsection{Low order shaping parameters}\label{sub:shaping}
\begin{figure}[tbp]
\includegraphics[width=0.47\columnwidth]{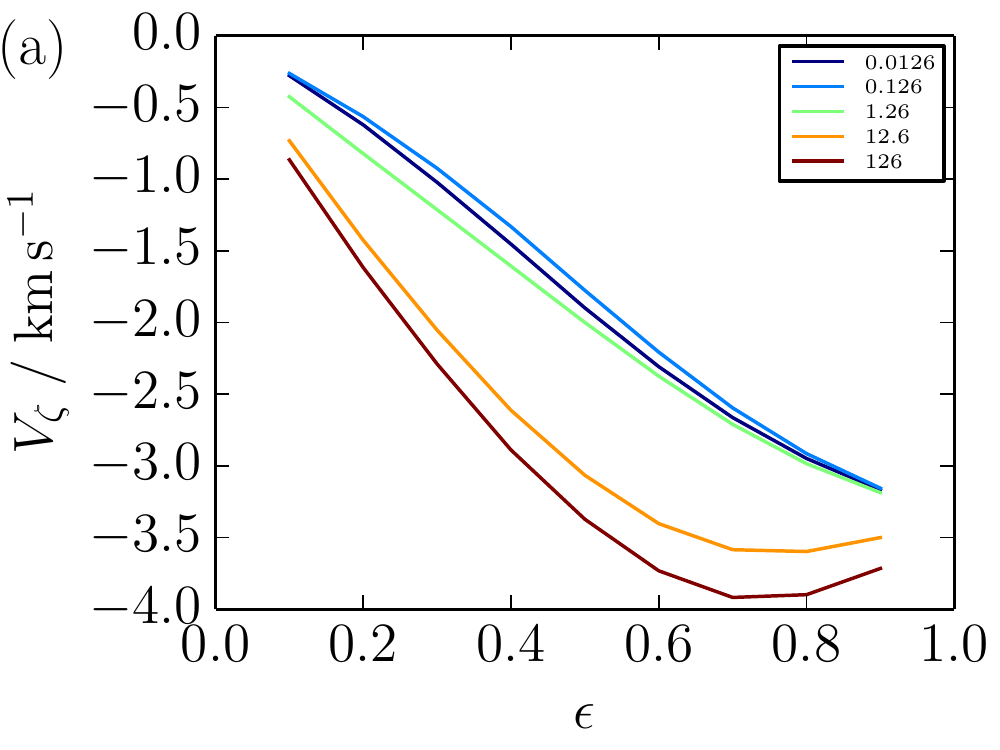}\hfill{}\includegraphics[width=0.47\columnwidth]{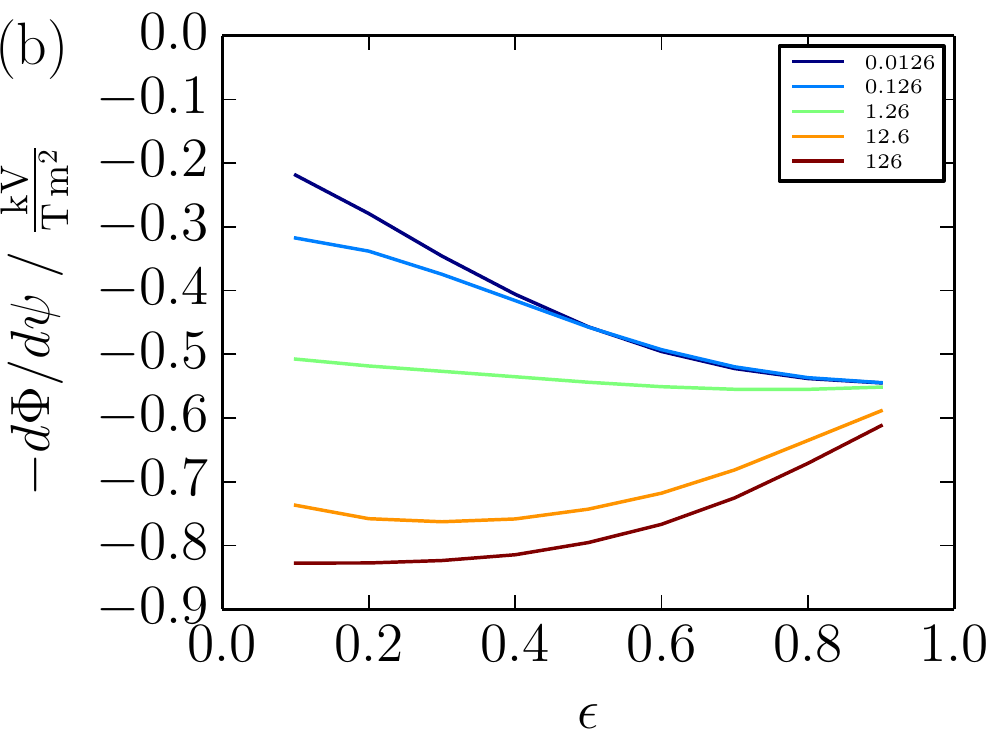}\\
\includegraphics[width=0.47\columnwidth]{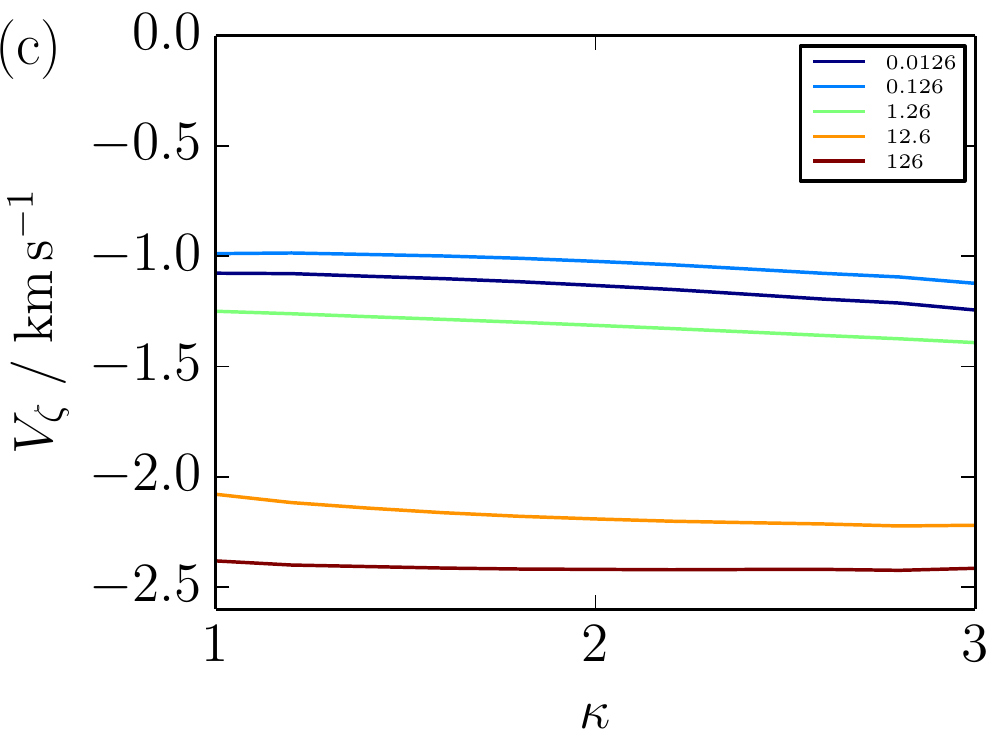}\hfill{}\includegraphics[width=0.47\columnwidth]{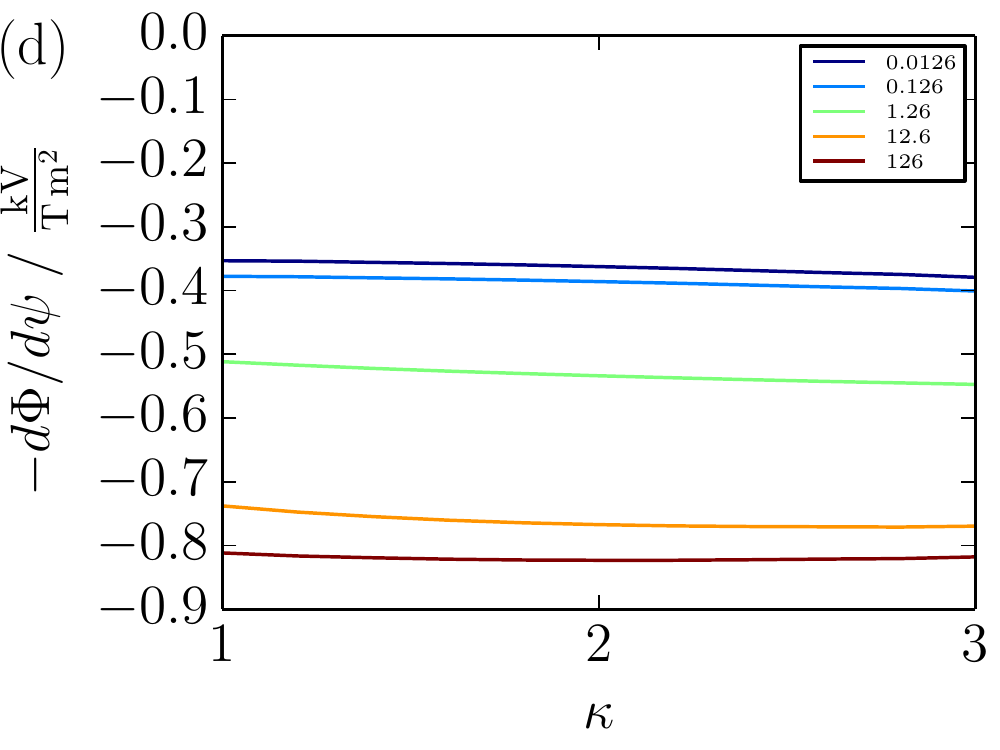}

\caption{Response of the toroidal rotation $\V$ and electric field
  $-d\Phi_0/d\psi$ at the outboard midplane to varying inverse aspect ratio
  $\epsilon$ (a,b) and elongation $\kappa$ (c,d) for several
  collisionalities, with neutrals localized at the X-point. The
  shaping parameters not being scanned are kept at the baseline
  values.
  Colours blue to red again correspond to low to high collisionalities
  as shown in the legend.
  %
  $d\Ti/d\psi$ is held constant at the value corresponding to a
  gradient length scale of 10 cm in the baseline geometry.
  \label{fig:epsilon-kappa}}
\end{figure}
In Ref.~\citep{Omotani2016} we studied the effect on the electric
field and rotation of varying the shaping parameters while keeping the
neutrals localized at $\tn=\theta_\mathrm{X}$, where
$\theta_\mathrm{X}$ is the poloidal angle of the X-point, representing
a divertor recycling source or private flux region fuelling. We
concluded that the shaping parameters most relevant for machine
operation, triangularity and X-point position, only affect the
electric field and rotation by changing the major radius where the
neutrals are localized, for a fixed collisionality.  Here we show that
the same is true for the elongation $\kappa$, as long as we hold fixed
the temperature gradient in poloidal flux space $d\Ti/d\psi$ rather
than the radial gradient $d\Ti/dr$.
Figs.~\ref{fig:epsilon-kappa}(c,d) show that varying $\kappa$ does not
change the radial electric field or rotation, as expected since the
major radius of the X-point does not vary with $\kappa$.

In contrast the response to varying the inverse aspect ratio
$\epsilon$ does not follow the same trend with the  major radius where
the neutrals are localized. In Ref.~\citep{Omotani2016} we found that
the electric field increased with $\Rn$ at high collisionality and
decreased at low collisionality. The radius of the X-point decreases
with increasing $\epsilon$, so Fig.~\ref{fig:epsilon-kappa}(b) shows
an opposite trend. In the unit aspect ratio limit, the mirror force on
the inboard side becomes large and suppresses poloidal flow, so that
the contributions parallel to $\boldsymbol B$ in (\ref{eq:V}) and
(\ref{eq:q}) vanish in this limit. This leaves only the rigid rotation
parts of the flow and heat flux, which are independent of
collisionality. The electric field then approaches
$-\,d\Phi_0/d\psi=(2/e)\,d\Ti/d\psi\approx
-0.547\;\mathrm{kV\,T^{-1}\,m^{-2}}$ (recall that $d\Ni/d\psi=0$ here)
and the outboard midplane toroidal rotation approaches
$\V=(R/e)d\Ti/d\psi\approx -3.39\;\mathrm{km\,s^{-1}}$, as can be seen
in Figs.~\ref{fig:epsilon-kappa}(a,b). In the large aspect ratio limit
$\epsilon\rightarrow 0$, $f_\mathrm{t}\rightarrow 0$ so $l\rightarrow
1$ in the banana regime (recall that $l=1$ generally in the
Pfirsch-Schl\"uter regime) while $\left\langle
B^2\right\rangle\rightarrow I^2/R^2$ so we see from (\ref{eq:q}) that
$q_\mathrm{i,\zeta}\rightarrow 0$. As discussed in section
\ref{sub:Interpretation} the heat flux gives rise to the intrinsic
momentum flux that drives rotation, Eq.~(\ref{eq:approx-flux}), and so
the rotation also vanishes $\V\rightarrow 0$ in this
limit.\footnote{Again, we ought to account for the third Legendre
harmonic component as well as the heat flux, but the accurate banana
regime result (\ref{eq:Vzeta_banana}) shows that the rotation indeed
vanishes, since $f_2\rightarrow 1$ while $R\approx\Rn\approx R_0$ and
$\sqrt{\left\langle B^2\right\rangle}\approx\Bn\approx B_0$.}

%

\subsection{Impurity rotation}

\begin{figure}
\includegraphics[width=0.47\columnwidth]{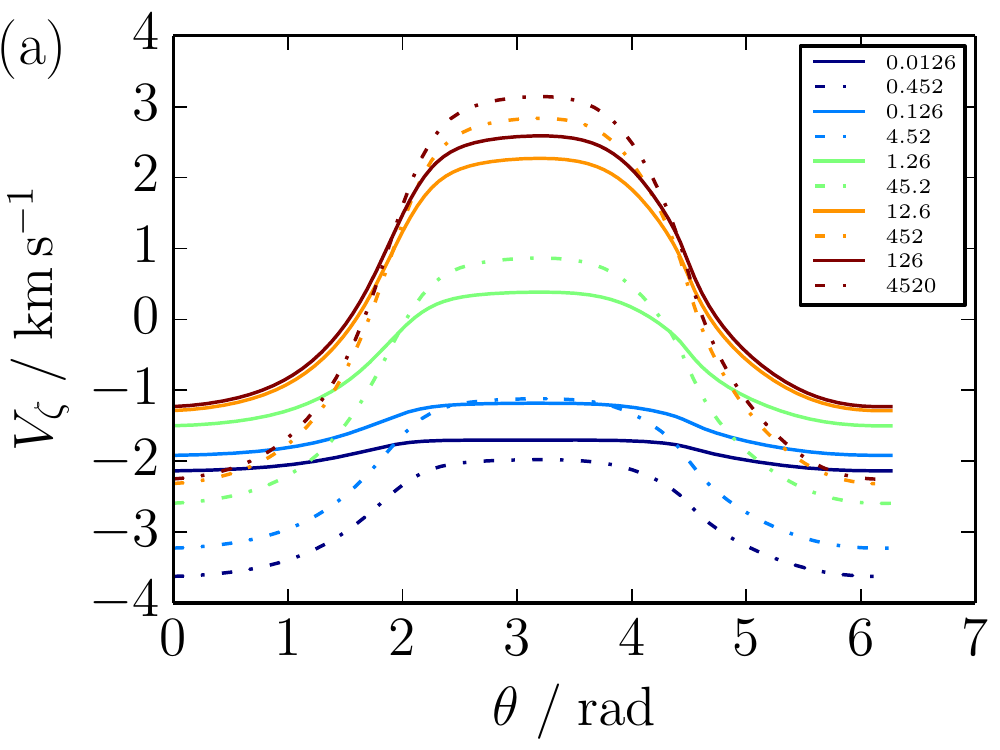}\hfill{}\includegraphics[width=0.47\columnwidth]{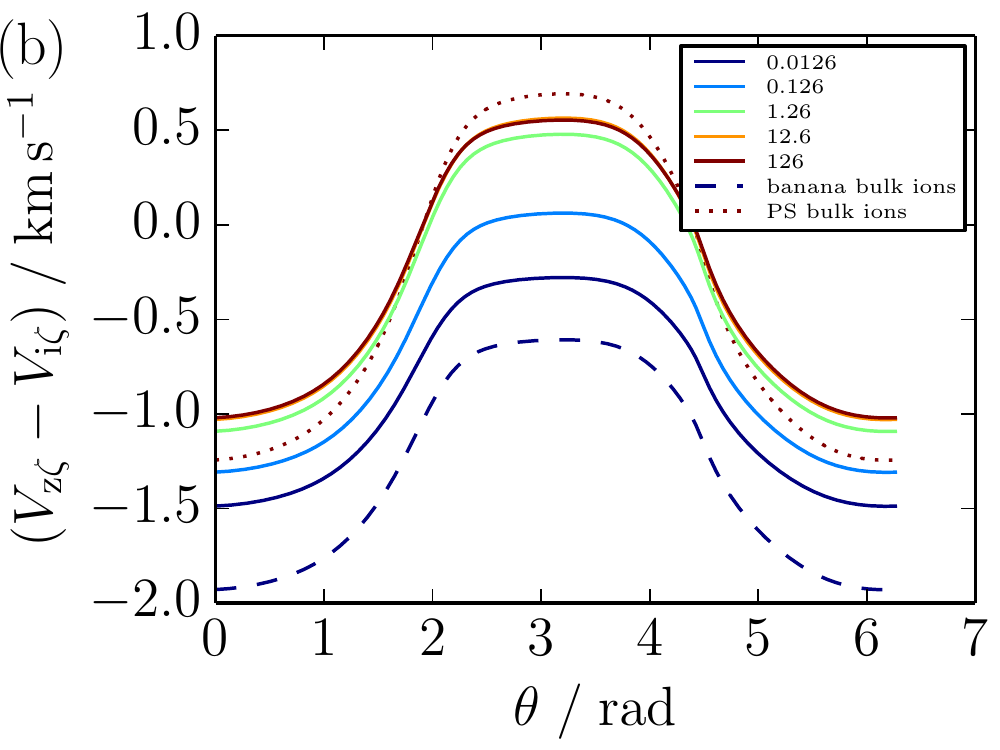}

\caption{Poloidal variation of (a) the toroidal velocity of bulk ions
  (solid) and trace, fully ionized carbon impurity (dash-dotted) for
  neutrals localized at the outboard midplane $\tn=0$ and (b) the
  difference between carbon and bulk ion velocities (solid), compared
  to analytical limits with collisional impurity and collisional
  (dotted) or banana regime (dashed) bulk ions. Colours from blue to
  red correspond to low to high collisionality $\hat\nu$, as shown in
  the legend. In (b), only the bulk ion collisionality is quoted.
  The ion temperature scale length is again taken to be 10 cm to
  produce explicit values. $\theta=0$ is the outboard midplane.
  \label{fig:impurity}}
\end{figure}


Tokamak rotation is most often measured by charge exchange
recombination spectroscopy \citep{Isler1994}. Due to the high
background of deuterium $D_\alpha$ radiation, an impurity species is
most often used for the measurement. However, it is well known both
theoretically \citep{Kim1991} and experimentally \citep{Kim} that
impurity and bulk ion rotation may differ. For a trace impurity
species, we may assume that the presence of the impurity does not
affect the momentum transport through neutrals, and by including the
impurity in our \textsc{perfect} simulations we can calculate the
impurity rotation along with that of the main ions, for arbitrary
collisionality of either species. The result is shown
in Fig.~\ref{fig:impurity}(a) for a trace, fully ionized carbon
impurity with density $n_\mathrm{z}\approx 0.018\Ni$ where the
neutrals are localized at the outboard midplane.
Ref.~\citep{Catto2006} gives a convenient form in their Eq.~(15) for
the velocity difference between bulk ions and a collisional impurity,
based on the results of Refs.~\citep{HirshmanSigmar81,Kim1991}. Here
we neglect all $\mathcal{O}(Z^{-1})$ terms for consistency, where $Z$
is the atomic number of the impurity.  These analytical results for
collisional and mixed collisionality (banana bulk ion and
Pfirsch-Schl\"uter impurity) limits are compared to the simulation
results in Fig.~\ref{fig:impurity}(b). The trends agree well, the
$\sim 30\%$ discrepancy is consistent with $\mathcal{O}(Z^{-1})$
errors for carbon and the approximate collision operators used
analytically. The analytical results show that the velocity difference
is independent of the radial electric field, and therefore independent
of the location of the neutrals in our framework, and indeed
Fig.~\ref{fig:impurity}(b) is identical for simulations with different
neutral locations. The quantitative differences highlight the
importance of accurate numerical results for the interpretation of
experimental impurity rotation measurements.

%

\section{Conclusions}

By considering the transport of toroidal momentum through
charge-exchanging neutrals, we calculate the radial electric field and
intrinsic rotation in a tokamak without external torque, where the
flows are subsonic. We have compared the results of our numerical
framework first presented in Ref.~\citep{Omotani2016} to the
analytical limits of Ref.~\citep{Helander2003} (with more accurate
collisional flow coefficients) finding good agreement in both
collisional and collisionless limits. We have extended the
investigation of Ref.~\citep{Omotani2016} to lower order shaping
parameters, finding that the elongation $\kappa$ has little effect on
the electric field and rotation due to neutrals. In contrast, varying
the inverse aspect ratio $\epsilon$ has more fundamental effects, for
instance changing the mirror forces, resulting in trends in electric
field and rotation that are not captured just by the collisionality
and the major radius where the neutrals are localized, $\Rn$. Finally,
in order to facilitate comparisons to experimental measurements which
usually obtain impurity rotation by charge exchange recombination
spectroscopy, we have demonstrated that impurities can be included in
the framework.  Here we neglect their interactions with the neutrals,
so impurities may be present only in trace quantities, but arbitrary
collisionality of either ion species can be treated.

\section*{Acknowledgements}
The authors are grateful to Matt Landreman for advice and help with
the \textsc{perfect} code and to Stuart Henderson for assistance with
the ADAS database. This work was supported by the Framework
grant for Strategic Energy Research (Dnr.~2014-5392) and the
International Career Grant (Dnr.~330-2014-6313) from
Vetenskapsr{\aa}det.

\bibliographystyle{iopart-num}
\bibliography{references}

\end{document}